\documentclass[table,xcdraw,fleqn,10pt]{wlscirep}

\usepackage[utf8]{inputenc}
\usepackage[T1]{fontenc}
\usepackage{comment}
\usepackage{lineno}
\usepackage{setspace}
\usepackage{colortbl}
\usepackage{xcolor}
\usepackage{placeins}    
\usepackage{float}       
\usepackage{tikz}
\usepackage[edges]{forest}
\usepackage{booktabs}    
\usepackage{multirow}    
\usepackage{tabularx}    
\usepackage{caption}
\usepackage{subcaption}
\usepackage{graphicx}
\usepackage[normalem]{ulem} 
\usepackage{tikz}
\usepackage{pgfplots}
\pgfplotsset{compat=1.18}
\usepgfplotslibrary{colormaps}
\usepackage{subcaption}
\definecolor{foldercolor}{RGB}{182,131,252}

\tikzset{
    pics/folder/.style={
        code={%
            \node[inner sep=0pt, minimum size=#1](-foldericon){};
            \node[folder style, 
                  inner sep=0pt, 
                  minimum width=0.3*#1, 
                  minimum height=0.6*#1, 
                  above right, 
                  xshift=0.05*#1] at (-foldericon.west){};
            \node[folder style, 
                  inner sep=0pt, 
                  minimum size=#1] at (-foldericon.center){};
        }
    },
    pics/folder/.default={20pt},
    folder style/.style={
        draw=foldercolor!80!black,
        top color=foldercolor!40,
        bottom color=foldercolor
    }
}

\forestset{
    is file/.style={
        edge path'/.expanded={%
            ([xshift=\forestregister{folder indent}]!u.parent anchor) |- (.child anchor)
        },
        inner sep=1pt
    },
    this folder size/.style={
        edge path'/.expanded={%
            ([xshift=\forestregister{folder indent}]!u.parent anchor) |- (.child anchor) 
            pic[solid]{folder=#1}
        }, 
        inner ysep=0.6*#1
    },
    folder tree indent/.style={
        before computing xy={l=#1}
    },
    folder icons/.style={
        folder, 
        this folder size=#1, 
        folder tree indent=3*#1
    },
    folder icons/.default={12pt},
}

\definecolor{myblue}{RGB}{220,230,241}
\definecolor{mygreen}{RGB}{230,240,230}
\definecolor{mygrey}{RGB}{245,245,245}
\definecolor{myheader}{RGB}{240,240,240}

\definecolor{section1}{RGB}{240, 248, 255} 
\definecolor{section2}{RGB}{240, 255, 240} 
\definecolor{section3}{RGB}{255, 240, 245} 
\definecolor{section4}{RGB}{245, 245, 220} 

\newcommand{\Vline}{\vrule width 0.2pt}
\renewcommand{\arraystretch}{1.4} 

\title{The iToBoS dataset: skin region images extracted from 3D total body photographs for lesion detection}

\author[1,*]{Anup Saha}
\author[1]{Joseph Adeola}
\author[2]{Nuria Ferrera}
\author[3]{Adam Mothershaw}
\author[10]{Gisele Rezze}
\author[4]{S\'eraphin Gaborit}
\author[5]{Brian D'Alessandro}
\author[6]{James Hudson}
\author[7]{Gyula Szab\'o}
\author[7]{Balazs Pataki}
\author[1]{Hayat Rajani}
\author[1]{Sana Nazari}
\author[1]{Hassan Hayat}
\author[3]{Clare Primiero}
\author[3,9]{H. Peter Soyer}
\author[2,8]{Josep Malvehy}
\author[1]{Rafael Garcia}

\affil[1]{Computer Vision and Robotics Research Institute, University of Girona, Girona, Spain}

\affil[2]{Dermatology Department, Hospital Cl\'inic Barcelona, Universitat de Barcelona, Barcelona, Spain}

\affil[3]{Frazer Institute, The University of Queensland, Dermatology Research Center, Brisbane, Australia}

\affil[4]{ISAHIT, Paris, France}

\affil[5]{Canfield Scientific, Inc., Parsippany, New Jersey, USA}

\affil[6]{V7, London, United Kingdom}

\affil[7]{HUN-REN Institute for Computer Science and Control, Budapest, Hungary}

\affil[8]{CIBER de Enfermedades Raras, Instituto de Salud Carlos III, Barcelona, Spain}

\affil[9]{Dermatology Department, Princess Alexandra Hospital, Brisbane, QLD, Australia}

\affil[10]{Dermatology Department, University of Trieste, Trieste, Italy}

\affil[*]{corresponding author: A. Saha (anup.saha@udg.edu)}

\begin{abstract}
Artificial intelligence has significantly advanced skin cancer diagnosis by enabling rapid and accurate detection of malignant lesions. In this domain, most publicly available image datasets consist of single, isolated skin lesions positioned at the center of the image. While these lesion-centric datasets have been fundamental for developing diagnostic algorithms, they lack the context of the surrounding skin, which is critical for improving lesion detection. The iToBoS dataset was created to address this challenge. It includes 16,954 images of skin regions from 100 participants, captured using 3D total body photography. Each image roughly corresponds to a $7 \times 9$ cm section of skin with all suspicious lesions annotated using bounding boxes. Additionally, the dataset provides metadata such as anatomical location, age group, and sun damage score for each image. This dataset aims to facilitate training and benchmarking of algorithms, with the goal of enabling early detection of skin cancer and deployment of this technology in non-clinical environments.
\end{abstract}

\begin{document}

\flushbottom
\maketitle

\thispagestyle{empty}



\section*{Background \& Summary}
Skin cancer is the most prevalent form of cancer globally, with Melanoma (MEL), Basal Cell Carcinoma (BCC), and Squamous Cell Carcinoma (SCC) representing the most common malignancies. These cancers have reached epidemic proportions, underscoring the critical need for effective diagnostic and monitoring solutions \cite{Garbe_2024}. Traditionally, the diagnosis of skin cancer has relied heavily on the expertise of dermatologists and the use of dermatoscopy. Dermatoscopy is a non-invasive approach that uses a dermoscope to enhance the view of the sub-macroscopic structures in pigmented skin lesions, which vary widely across dermatological conditions \cite{conforti2020dermoscopy,dinnes2018visual}. As a result, they provide critical cues, not only for the visual examination of suspicious lesions by trained dermatologists but also for AI-based diagnostic tools \cite{zhang2023ecl, kim2023skin, Naqvi_2023, dildar2021skin, Saeed_2021, ali2020towards, Xu_2018, Namozov_2018} that aid in differentiating between malignant and benign skin conditions. Consequently, most publicly available skin lesion datasets for training AI models are dominated by dermoscopic images \cite{Codella_2018, Tschandl_2018, Hern_ndez_P_rez_2024, Rotemberg_2021}.

However, while dermoscopy has resulted in improved diagnostic precision, its reliance on specialised equipment and skilled practitioners poses significant barriers to early detection and triage. In such contexts, computer-aided diagnosis systems that utilise conventional camera images offer a more practical and scalable solution. These systems, designed to work with standard smartphone cameras, can be deployed in resource-constrained environments, enabling non-specialist healthcare providers\textemdash such as general practitioners or community health workers\textemdash to identify potentially malignant lesions. 3D total-body photography (3D-TBP) is increasingly gaining recognition as a foundational tool in this regard \cite{janda2018using,rayner2018clinical,primiero2019evaluation}. Such imaging systems use a coordinated setup of cameras to capture comprehensive, high-resolution views of almost the entire skin surface in a single session, providing a detailed and holistic representation of a patient’s skin. This not only provides a broader perspective for lesion analysis by incorporating the surrounding skin area that might be otherwise excluded when assessing a single localised lesion, but it can also facilitate longitudinal tracking of lesions over time to monitor disease progression or treatment response. 

Canfield's VECTRA WB360 is a state-of-the-art 3D-TBP system that provides automatic lesion detection capabilities. Recently, the International Skin Imaging Collaboration (ISIC) used this system to compile a lesion classification dataset \cite{ISIC2024} consisting of lesion-centric crops extracted from 3D-TBP scans as part of their lesion classification challenge, ISIC 2024, which garnered huge response. However, the system's automated lesion detection occasionally introduced false positives such as knuckles, nipples, bellybuttons, and tattoos. All images were manually reviewed to remove erroneously classified samples and refine the dataset \cite{ISIC2024}. Additionally, the system was noted to over-identify solar damage lesions. On the other hand, there remains a lack of publicly available datasets to train AI models to improve on this task. As such, we present a novel dataset comprising 16,954 high-resolution images from two different sites: Hospital Clinic Barcelona, Spain and The University of Queensland, Brisbane, Australia. The contribution of every site includes 51 patients from Barcelona and 49 from Brisbane. These images are tiles of patients' skin surface captured using the VECTRA WB360 3D-TBP system (Canfield Scientific). The dataset encompasses images from diverse anatomical locations, including the torso, arms and legs, but excluding the face to preserve patient anonymisation. Each image is annotated with bounding boxes that delineate skin lesions, providing a rich dataset for training AI models for lesion detection. By capturing a diverse set of lesions across different anatomical regions and patient demographics, the dataset aims to reflect the variability seen in real-world clinical practice. The images are also accompanied by metadata such as the patient's age, anatomical region and sun damage score corresponding to the image. These parameters were carefully chosen to provide additional context for the AI models, helping them distinguish between lesions and healthy skin by more effectively accounting for demographic and environmental factors that influence these conditions.

This dataset is also a key component of the iToBoS-2024 Skin Lesion Detection Challenge, hosted on Kaggle. The objective of this competition is to develop state-of-the-art machine learning techniques for detecting multiple skin lesions in clinical images, with the overarching goal of advancing the timely diagnosis and treatment of skin cancer.

\section*{Methods}
\begin{figure}[t!]
    \centering
    \includegraphics[width=0.85\linewidth]{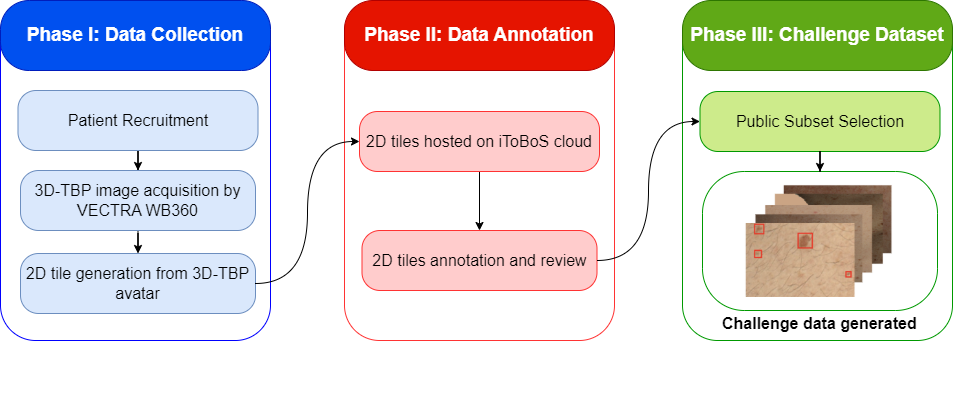}
    \caption{Three-phase methodology for dataset generation: (i) data collection through 3D-TBP imaging and 2D tile extraction, (ii) data annotation and expert review, and (iii) careful selection of a representative public subset to be released as the challenge dataset.}
    \label{methodimage}
\end{figure}

The process for generating the dataset involved three key phases: (i) data collection, (ii) data annotation, and (iii) public subset selection, as summarised in Figure \ref{methodimage}. Phase (i), data collection, began with patient recruitment at two clinical sites, followed by capturing 3D-TBP images using the VECTRA WB360 scanner and the extraction of 2D tiles from the 3D avatar. Phase (ii), data annotation, included hosting these tiles on the iToBoS cloud platform, annotation and quality control. Finally, phase (iii) involved the careful selection of a public subset for release, resulting in the challenge dataset. Each of these phases is elaborated in the following subsections.

\subsection*{Data Collection}

\subsubsection*{Patient Recruitment}
The study was conducted in two locations: Brisbane (Australia) and Barcelona (Spain). Participants in Australia were recruited via email between October 2022 and April 2023 and provided online consent for the use of existing 3D-TBP images captured during previous research visits, which took place from September 2016 to February 2020 using the VECTRA WB360 system. 

Enrolment for the Barcelona site began in January 2023 and was completed in March 2024. In Barcelona, individuals who either had a personal history of melanoma, a family history of melanoma, or exhibited atypical nevus syndrome were invited for a clinical evaluation at the Hospital Clinic Barcelona. The imaging at the Barcelona site was performed prospectively (with the consent of the participants) with the same VECTRA WB360 system. 

To ensure comprehensive data collection and tracking, each participant was assigned a unique identifier, with some participants undergoing multiple imaging sessions over time. Additionally, all participants completed detailed questionnaires capturing demographic, sun protective behaviour and phenotypic information, along with relevant medical history.



\subsubsection*{3D-TBP image acquisition system of VECTRA WB360}
Each participant underwent total body imaging using the VECTRA WB360 system. This imaging system was designed for comprehensive skin documentation through efficient, high-resolution capture. The system utilizes 92 fixed cameras (arranged in 46 stereo pairs) and xenon flashes to photograph the entire exposed body in a single capture. To ensure standardised image quality, participants were instructed to maintain a specific anatomical position while the system captured detailed images using both polarised and non-polarised lighting.
The captured images were processed by the VECTRA software to create a precise 3D avatar, enabling full 360-degree rotation for thorough examination of all body surfaces, including curved areas that are often challenging to assess with traditional 2D imaging \cite{Clare_P}. 
All imaging data was saved in DX2 format, a well-known file type used for various applications, including medical imaging and dermatological assessments. Each DX2 file contains the original 2D images, the reconstructed 3D avatar, and all associated metadata required for dermatological assessment.

\subsubsection*{2D tile generation from 3D-TBP avatar}
The process of extracting images corresponding to 2D tiles of the skin surface from 3D avatars was accomplished using \textit{WbTilingTool}, a specialised module within the \textit{VectraDBTool} developed by Canfield Scientific. This tool was designed to streamline and enhance image processing by automatically detecting the patient's head in 3D avatars and applying inpainting techniques to mask facial features, ensuring patient anonymity. Its robust functionality includes batch processing, enabling the simultaneous handling of multiple scans from various patients, significantly improving efficiency for large datasets.

\begin{figure}[b!]
    \centering
    \includegraphics[width=0.8\linewidth]{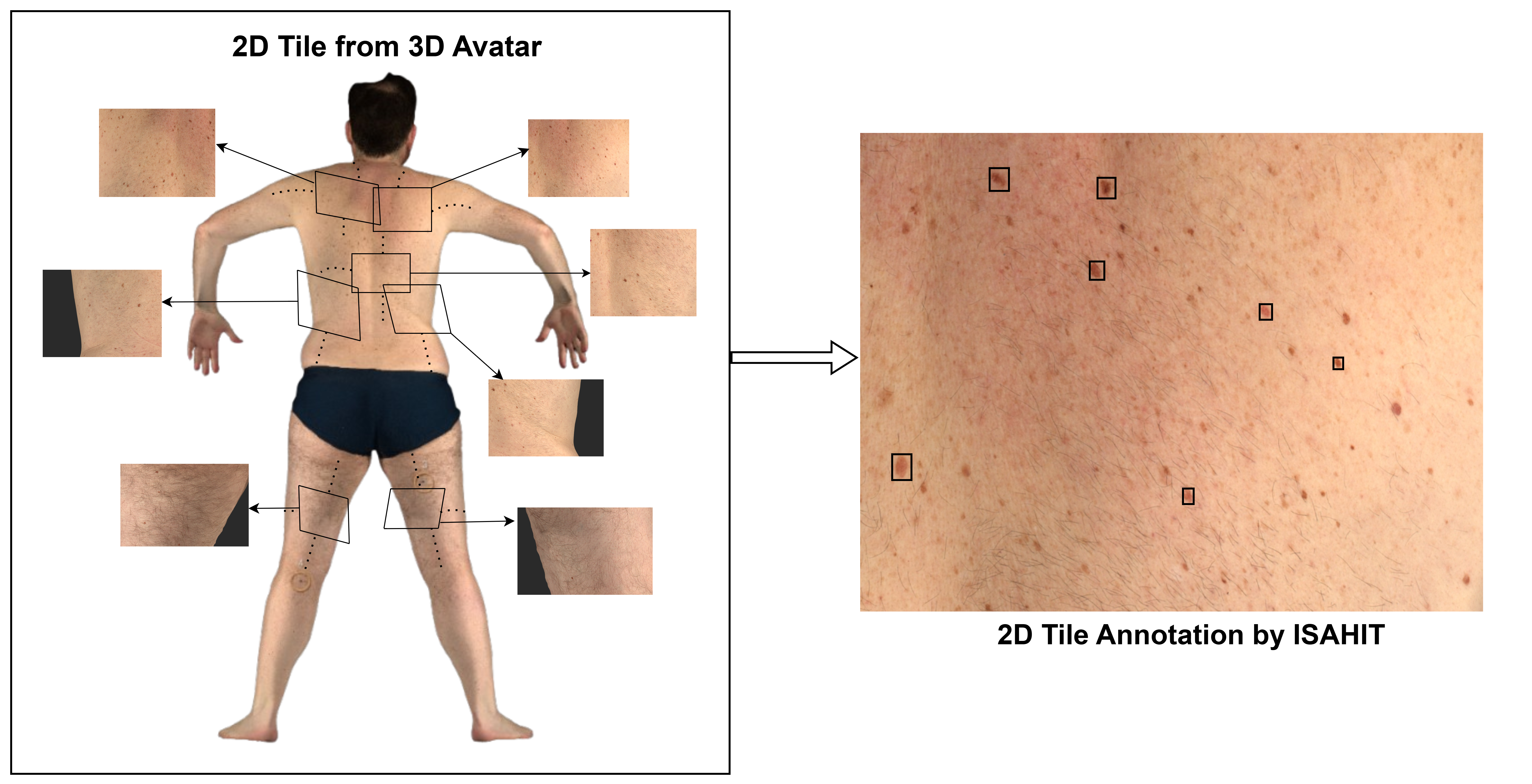}
    \caption{2D tiles generation from 3D-TBP: The iToBoS dataset consists of the patients' skin surface tiles generated from their 3D avatars. Subsequently, these images were annotated by ISAHIT and quality checked by clinicians on the V7 Darwin platform.}
    \label{tileimage}
\end{figure}

Figure \ref{tileimage} shows an example of a 2D tiles generated from a 3D avatar. Each avatar was split into tiles with an average dimension of 1012$\times$827 pixels $(px)$, with a $45 px$ overlap between adjacent tiles. This overlap was crucial for maintaining lesion visibility across tile boundaries, particularly important for lesions that might otherwise be split between tiles. Tiles at the edges and bottom of the avatar had smaller dimensions to accommodate the natural boundaries of the body surface. This variation in tile size occurred specifically at these boundary regions to ensure complete coverage of the avatar while avoiding truncation. Each tile was tagged with its corresponding anatomical region to preserve the anatomical context. The images were categorised into five anatomical regions plus an \textit{Unknown} category for ambiguous cases, as summarised in Table \ref{table:body-part-categorisation}. This step, along with the extraction of metadata such as age and sex-at-birth from questionnaires collected from each participant, was crucial for ensuring a balanced distribution of samples, particularly when selecting the subset of the dataset designated for public release, as detailed later. Representative examples of images from each anatomical region category are shown in Figure \ref{fig:body_parts}.
\begin{table}[h!]
\centering
\caption{Anatomical Region Categorisation for Image Classification.}
\begin{tabular}{|p{0.25\textwidth}|p{0.65\textwidth}|}
\hline
\rowcolor{myheader}
\textbf{Region} & \textbf{Description} \\ \hline
\rowcolor{section2} Torso &  Anterior and posterior regions of the trunk, excluding limbs \\ \hline
\rowcolor{section2} Left Arm & Hand, forearm, and upper arm up to the shoulder \\ \hline
\rowcolor{section2} Right Arm & Hand, forearm, and upper arm up to the shoulder \\ \hline
\rowcolor{section2} Left Leg & Foot, lower leg (shin and calf), and upper leg (thigh) \\ \hline
\rowcolor{section2} Right Leg & Foot, lower leg (shin and calf), and upper leg (thigh) \\ \hline
\rowcolor{section2} Unknown & Cases where exact body part was not identified or documented, which could involve unspecified areas or those outside the defined body parts in the dataset.  \\ \hline
\end{tabular}
\label{table:body-part-categorisation}
\end{table}


\newlength{\figurewidth} 
\setlength{\figurewidth}{0.2\textwidth}
\newlength{\imagewidth} 
\setlength{\imagewidth}{0.2\linewidth}
\newlength{\rowspacing} 
\setlength{\rowspacing}{0.05cm}

\begin{figure}[h!]
    \centering
    
    \begin{minipage}{\figurewidth}
        \centering
        \includegraphics[width=\imagewidth, keepaspectratio]{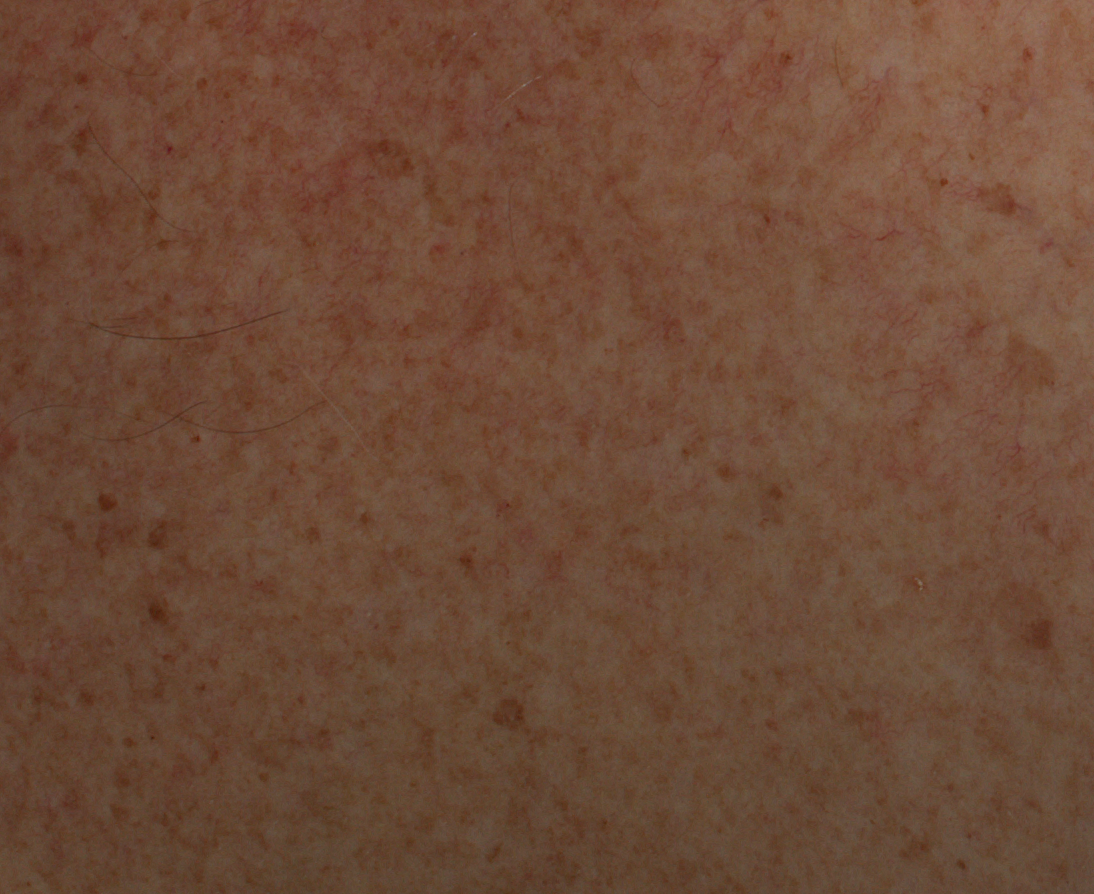}
        \caption*{Torso 1}
    \end{minipage}
    \hfill
    \begin{minipage}{\figurewidth}
        \centering
        \includegraphics[width=\imagewidth, keepaspectratio]{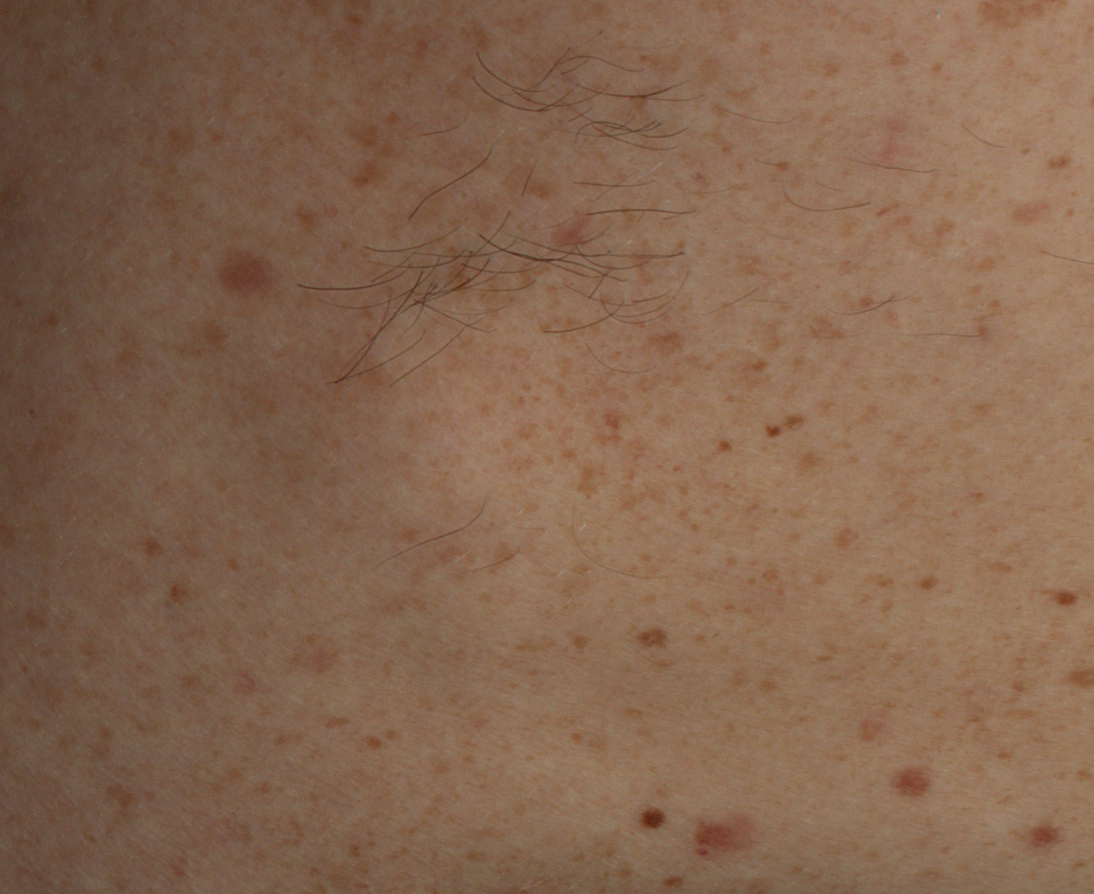}
        \caption*{Torso 2}
    \end{minipage}
    \hfill
    \begin{minipage}{\figurewidth}
        \centering
        \includegraphics[width=\imagewidth, keepaspectratio]{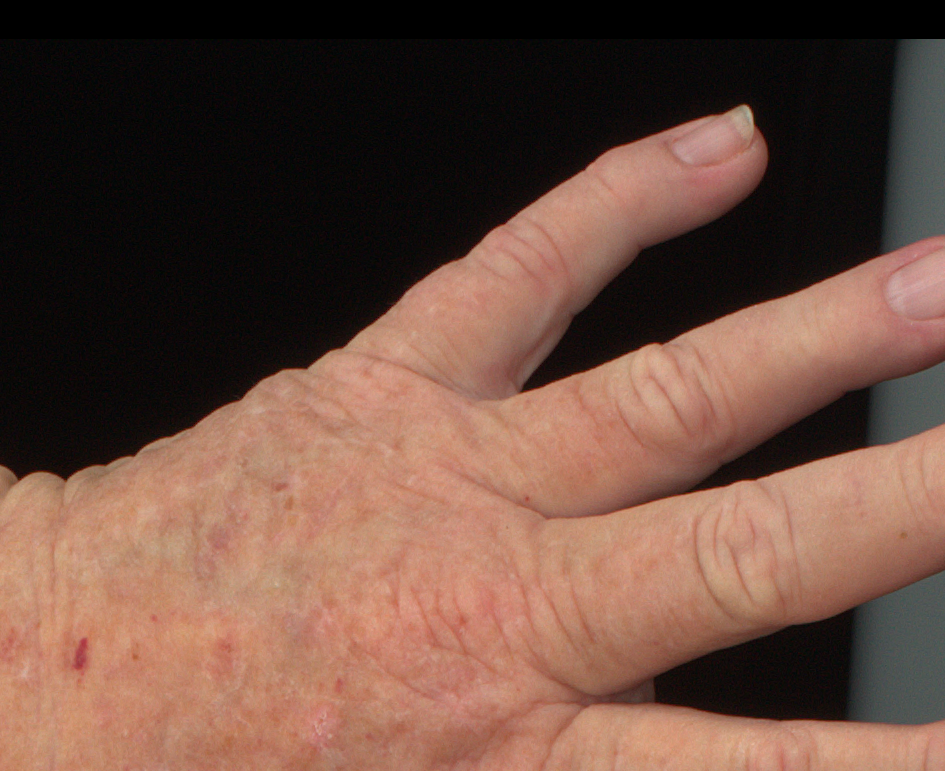}
        \caption*{Left Arm 1}
    \end{minipage}
    \hfill
    \begin{minipage}{\figurewidth}
        \centering
        \includegraphics[width=\imagewidth, keepaspectratio]{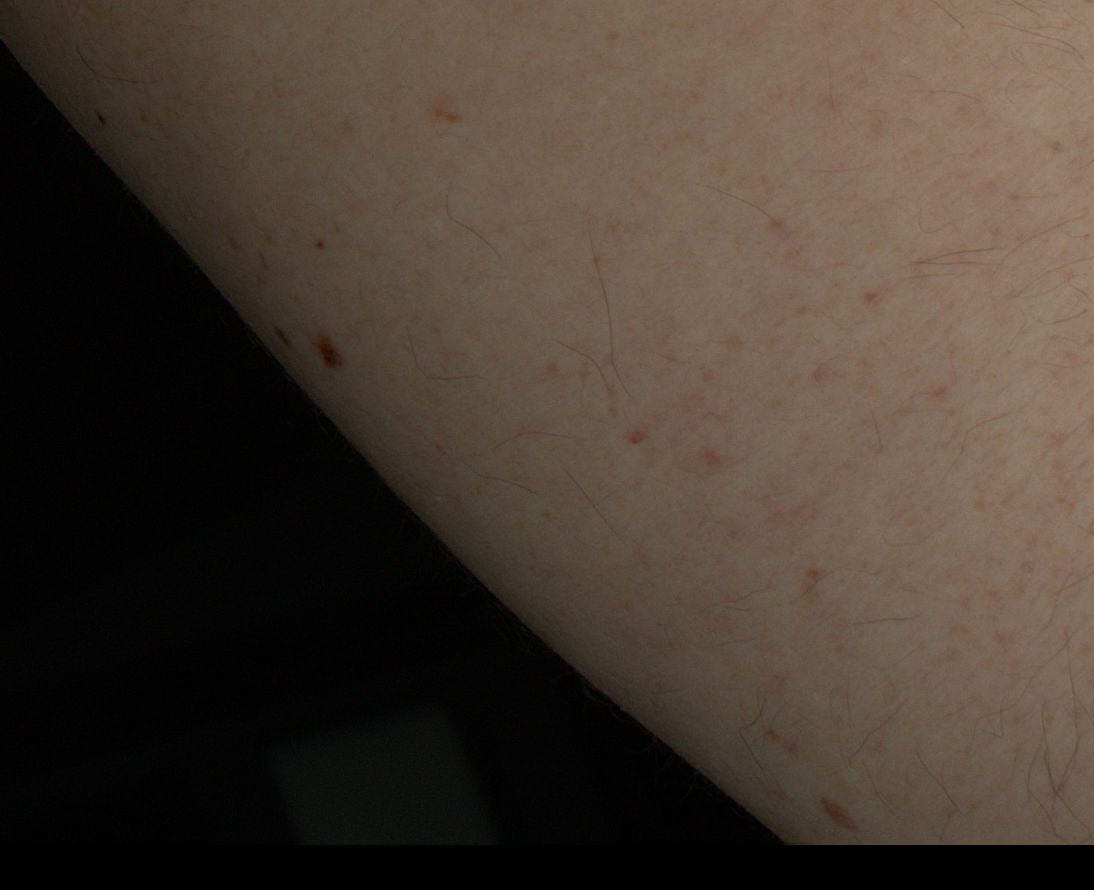}
        \caption*{Left Arm 2}
    \end{minipage}
    \vspace{\rowspacing} 
    
    \begin{minipage}{\figurewidth}
        \centering
        \includegraphics[width=\imagewidth, keepaspectratio]{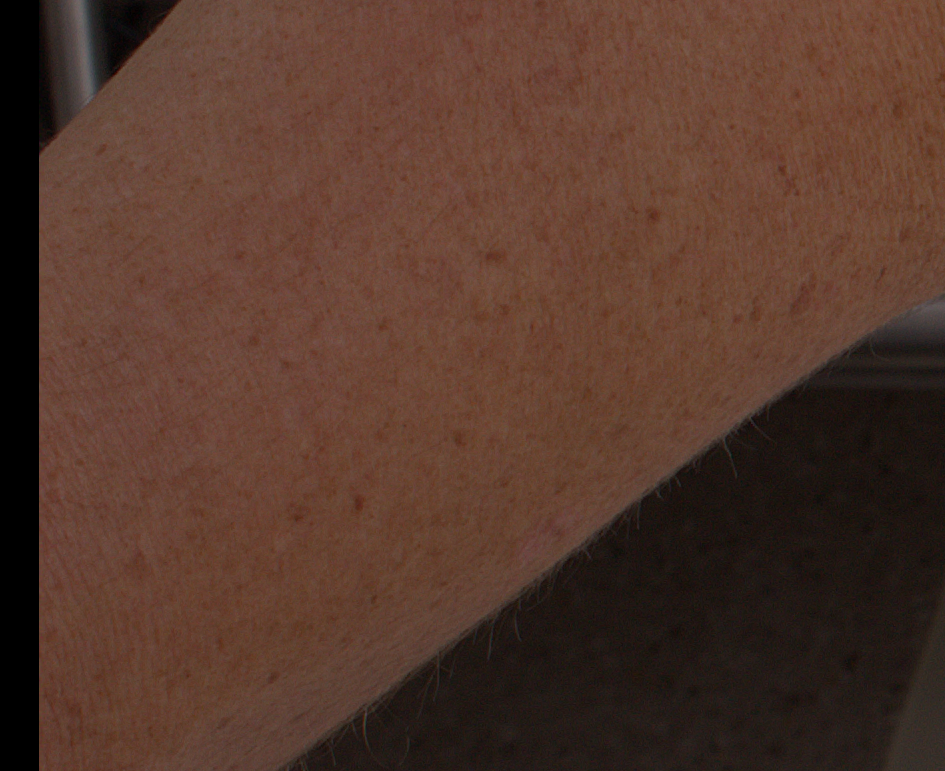}
        \caption*{Right Arm 1}
    \end{minipage}
    \hfill
    \begin{minipage}{\figurewidth}
        \centering
        \includegraphics[width=\imagewidth, keepaspectratio]{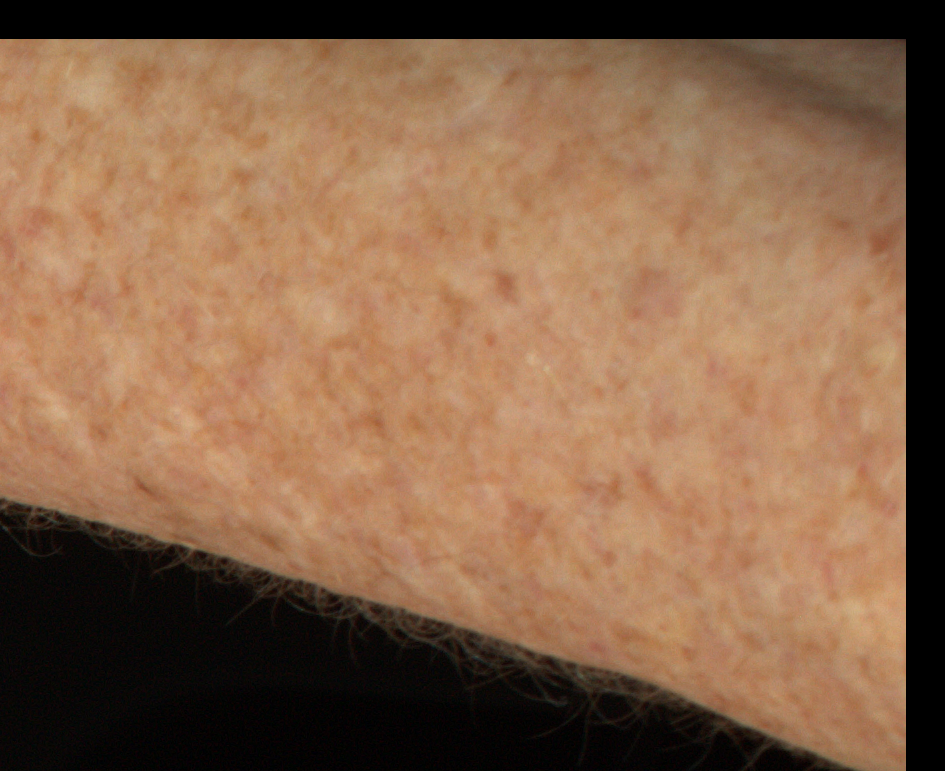}
        \caption*{Right Arm 2}
    \end{minipage}
    \hfill
    \begin{minipage}{\figurewidth}
        \centering
        \includegraphics[width=\imagewidth, keepaspectratio]{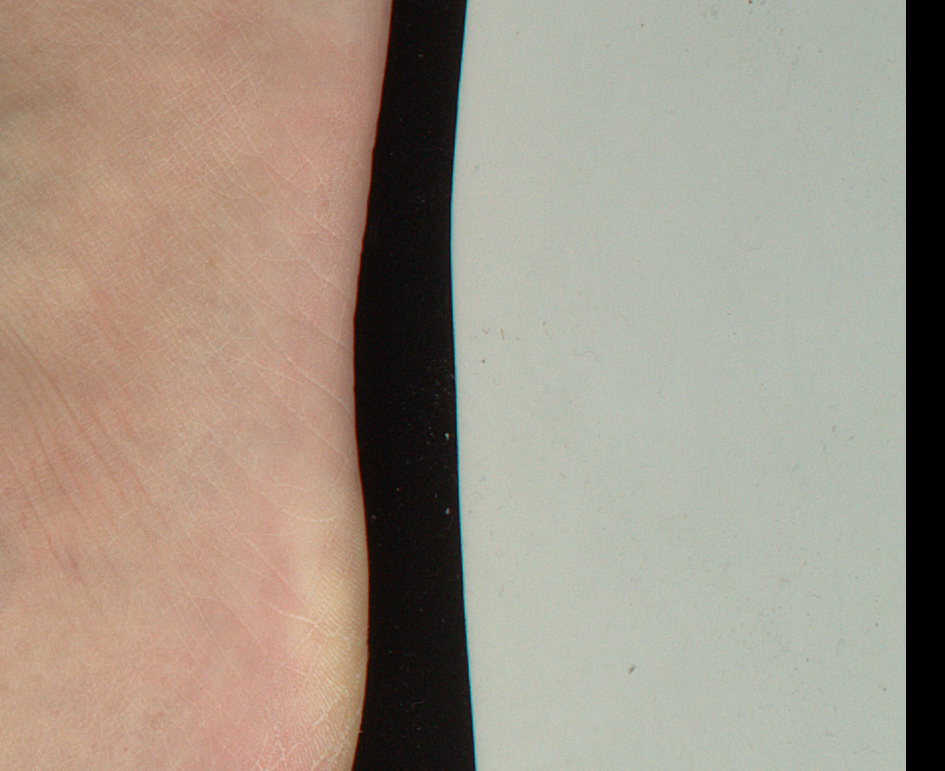}
        \caption*{Left Leg 1}
    \end{minipage}
    \hfill
    \begin{minipage}{\figurewidth}
        \centering
        \includegraphics[width=\imagewidth, keepaspectratio]{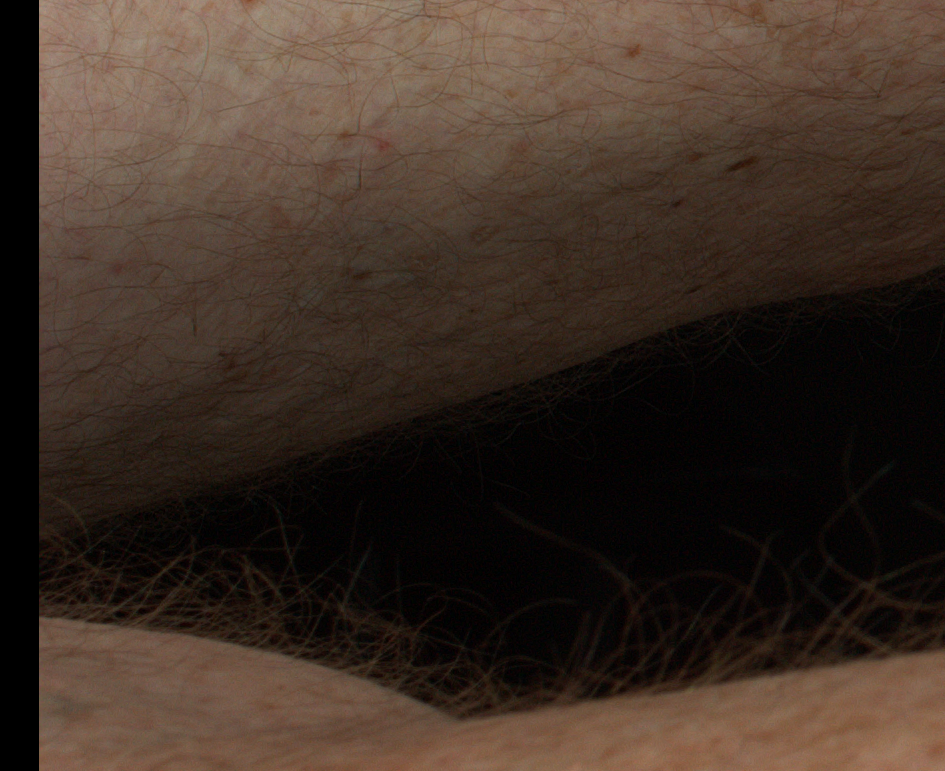}
        \caption*{Left Leg 2}
    \end{minipage}
    \vspace{\rowspacing} 

    \begin{minipage}{\figurewidth}
        \centering
        \includegraphics[width=\imagewidth, keepaspectratio]{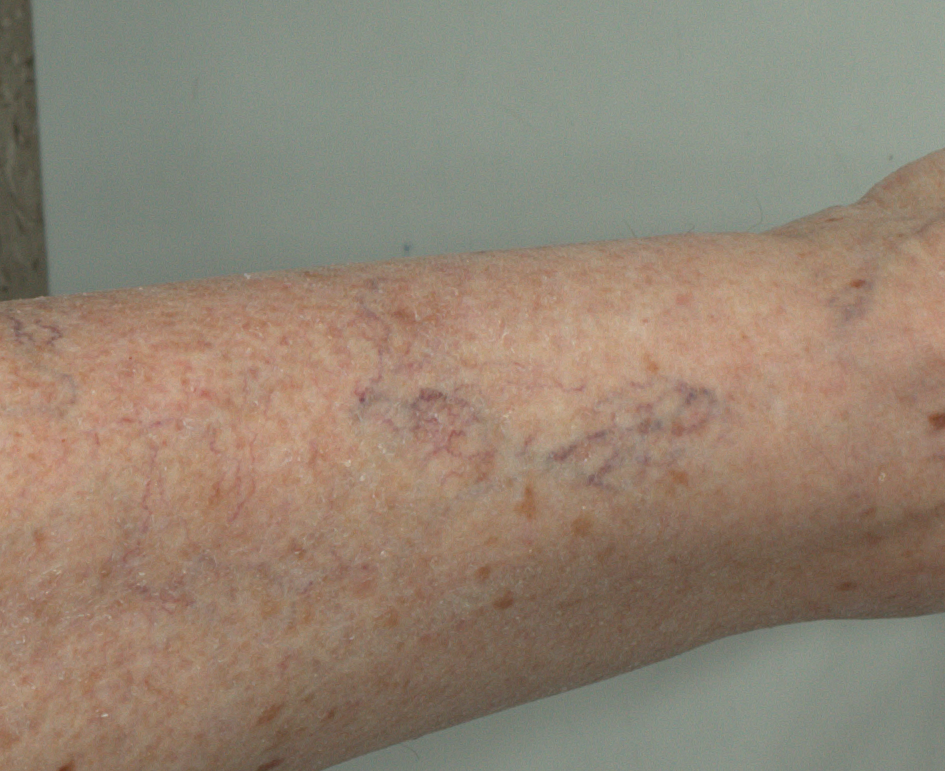}
        \caption*{Right Leg 1}
    \end{minipage}
    \hfill
    \begin{minipage}{\figurewidth}
        \centering
        \includegraphics[width=\imagewidth, keepaspectratio]{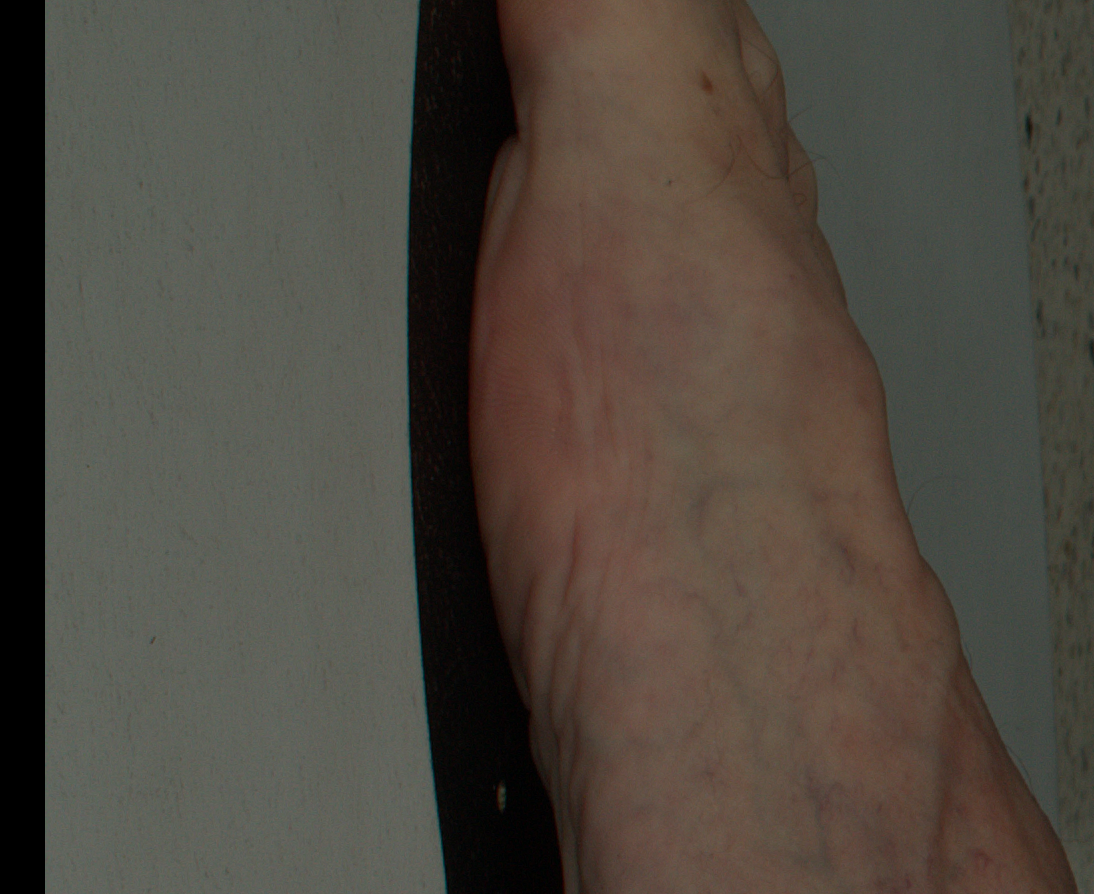}
        \caption*{Right Leg 2}
    \end{minipage}
    \hfill
    \begin{minipage}{\figurewidth}
        \centering
        \includegraphics[width=\imagewidth, keepaspectratio]{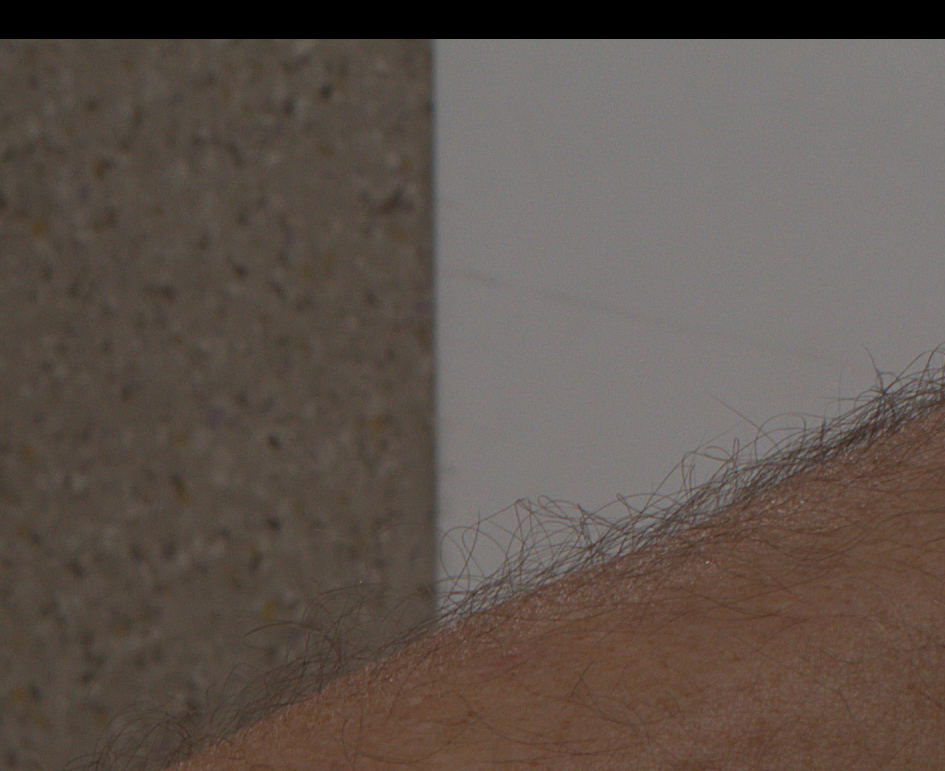}
        \caption*{Unknown1}
    \end{minipage}
    \hfill
    \begin{minipage}{\figurewidth}
        \centering
        \includegraphics[width=\imagewidth, keepaspectratio]{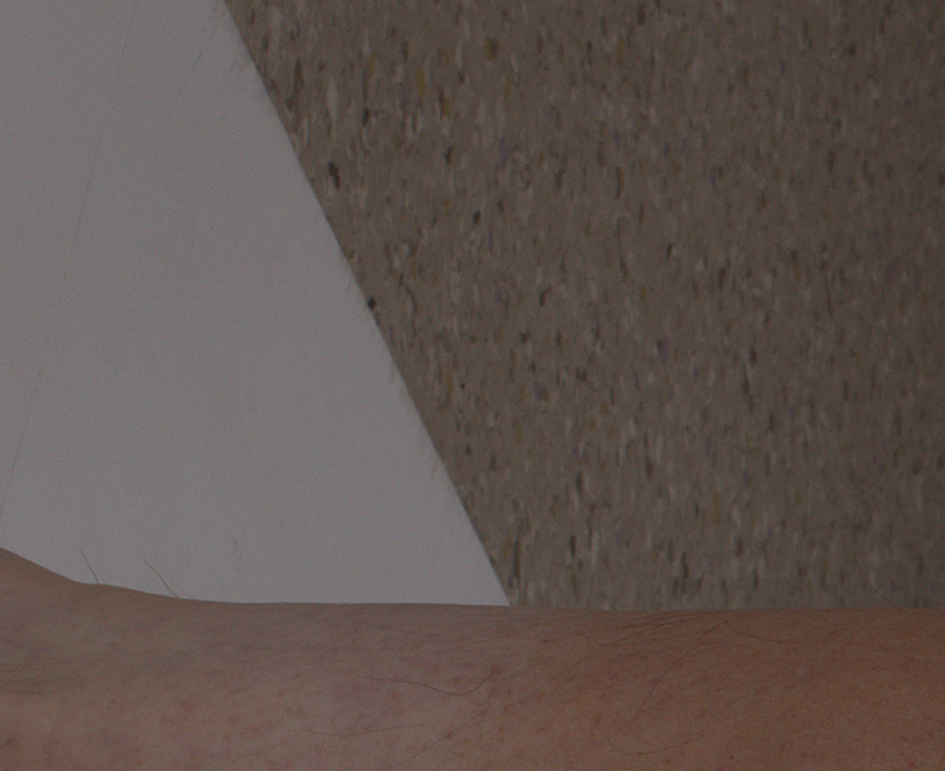}
        \caption*{Unknown2}
    \end{minipage}
    
    \caption{Sample 2D tiles illustrating the six anatomic region categories: torso (anterior and posterior), left arm, right arm, left leg, right leg, and unknown regions.}
    \label{fig:body_parts}
\end{figure}

The 3D spatial coordinates associated with each tile were also removed to enhance privacy. This critical step ensured that the tiles could not be used to reconstruct a 3D avatar of the patient, eliminating any potential risk of re-identification through spatial mapping. Additionally, the file names of the extracted 2D tiles were randomised, further reducing the risk of traceability or association with the original scans. As a further safeguard, visible marks, tattoos, and other distinguishing features in the tiles were meticulously masked out during the annotation stage. These steps ensured that patient identities were thoroughly protected, compliant with GDPR guidelines for protecting the privacy of the patients.

\subsection*{Data annotation}
The 2D tiles extracted from the 3D avatars were annotated by a team of 25 annotators coordinated by Isahit, a company specialised in annotation services across various domains, including medical and image-based data. The team, consisting of medical students, nurses and doctors, received specialised training to accurately identify and label skin lesions. At the beginning of the annotation process, the annotators were divided into two groups: \textit{labellers}, responsible for the initial annotations, and \textit{reviewers}, who ensured the quality of the annotations.

The \textit{labellers} began by assessing the level of sun damage on each tile, assigning a severity tag from 1 to 3, with level 1 indicating low and level 3 indicating high sun damage. Then, each lesion was outlined using a bounding box and its colour was annotated to enhance characterisation. Each tile was also tagged if it contained any tattoos or identifiable marks, thereby providing an additional layer of verification for challenging cases and enabling their subsequent annotation and masking. Finally, \textit{reviewers} manually checked each tile to verify the accuracy of the annotations. If a tile met the quality standards, it was marked as complete; otherwise, it was sent back to the \textit{labellers} for re-annotation.

\subsubsection*{2D tiles hosted on iToBoS cloud}
The iToBoS cloud, managed by the Computing and Automation Research Institute, HUN-REN SZTAKI, played a crucial role in the annotation process by taking on two key responsibilities: providing secure storage for the 2D tiles and managing the transfer of these data to and from the V7 Darwin annotation platform. This process was handled via a NextCloud service, supported by HUN-REN's cloud infrastructure. NextCloud allowed for seamless data uploads and management within a structured directory system, with strict authentication and authorisation protocols to ensure that only authorised personnel could access these files.

Once the files to be annotated were gathered in specific folders on the NextCloud platform, the upload scripts were used to create datasets in V7 Darwin platform. Then, Isahit's annotators used these datasets to perform the necessary annotations on the tiles. Therefore, dermatologists verified the Isahit annotation on V7 Darwin platform. After all datasets had been annotated and verified, the annotations were downloaded from V7 Darwin platform back into designated folders on NextCloud.

This workflow provided a flexible and asynchronous approach, facilitating the smooth transfer of primary data to NextCloud, as well as the generation and retrieval of derivative data, including the annotations, back into the system.

\subsubsection*{2D tiles annotation and review}
The V7 Darwin platform was used to annotate the tiles. This platform provided facilities for viewing the tiles and included a pre-defined set of tools, not only for polygonal annotation of lesions but also for assigning a sun damage score to each tile. The assessment of sun damage provided additional information for lesion analysis, as areas with higher sun damage typically presented a greater risk for developing skin lesions and influenced their appearance and characteristics. A circular "ruler" with a diameter of 2.5 mm was also provided with each tile to help the annotators determine the size of the lesions being annotated. In this dataset, a minimum threshold of 2.5 mm was established because, although smaller lesions may occasionally be melanomas, they are less common and often present significant diagnostic challenges \cite{Bono_2006}. In addition, V7 Darwin platform enabled the design and deployment of custom workflows. These workflows facilitated the assignment of lesion annotation and review tasks to Isahit's \textit{labellers} and \textit{reviewers}, followed by a final review by clinicians. The platform also enabled simultaneous operations by all participants in the workflow, allowing them to annotate lesions, add tags, accept or reject annotations, and provide feedback, thereby streamlining the annotation process and ensuring efficiency across tasks. The workflows were modular and were adjusted throughout the project to achieve the highest possible level of accuracy. For example, requirements for multiple reviewers in sequence were introduced as needed.

\begin{figure}
    \centering
    \includegraphics[width=\linewidth]{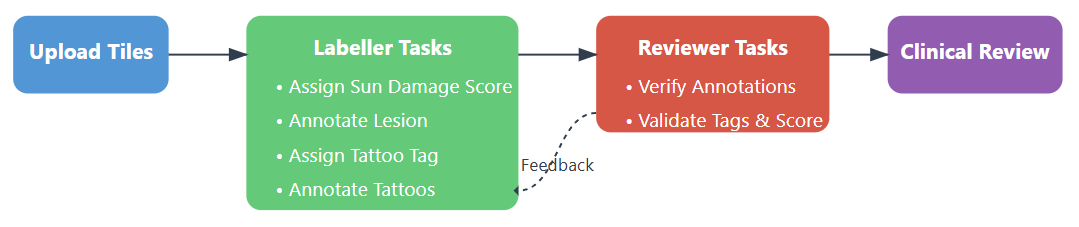}
    \caption{V7 Darwin platform annotation workflow illustrating the sequential process from tile upload to clinical review. Labellers perform sun damage scoring, tattoo tagging, and lesion annotation, while reviewers verify the completeness and accuracy of all labeller tasks. Dashed line represents the feedback loop for refinements.}
    \label{fig:v7-workflow}
\end{figure}

\begin{figure}[ht!]
    \centering
    \begin{minipage}{0.23\textwidth}
        \centering
        \includegraphics[width=\linewidth]{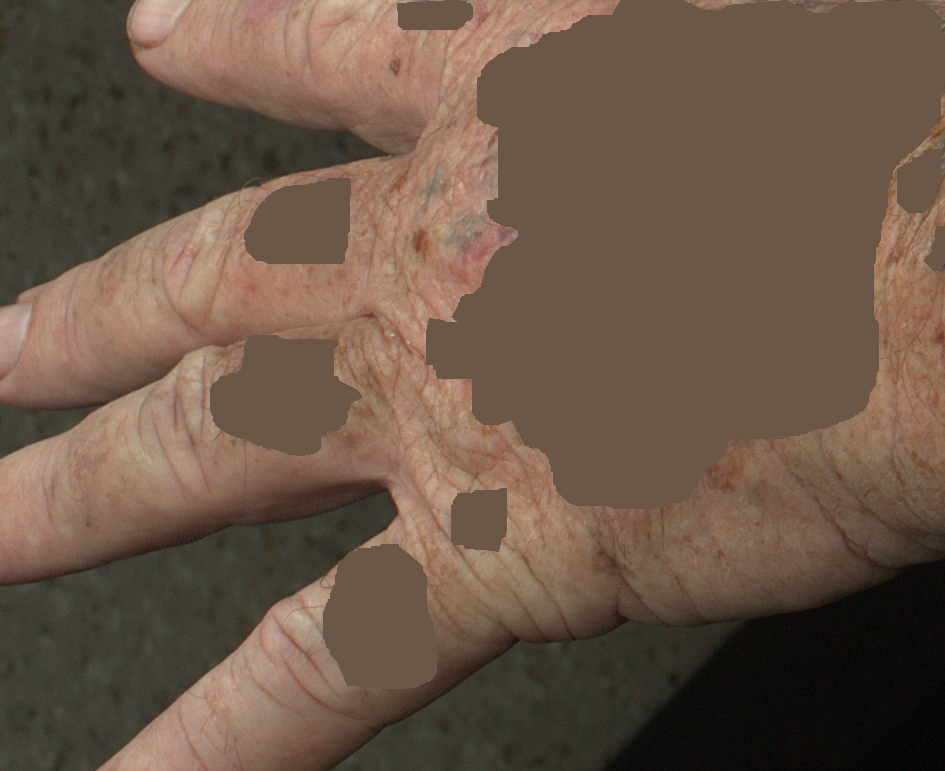}
    \end{minipage}
    \hfill
    \begin{minipage}{0.23\textwidth}
        \centering
        \includegraphics[width=\linewidth]{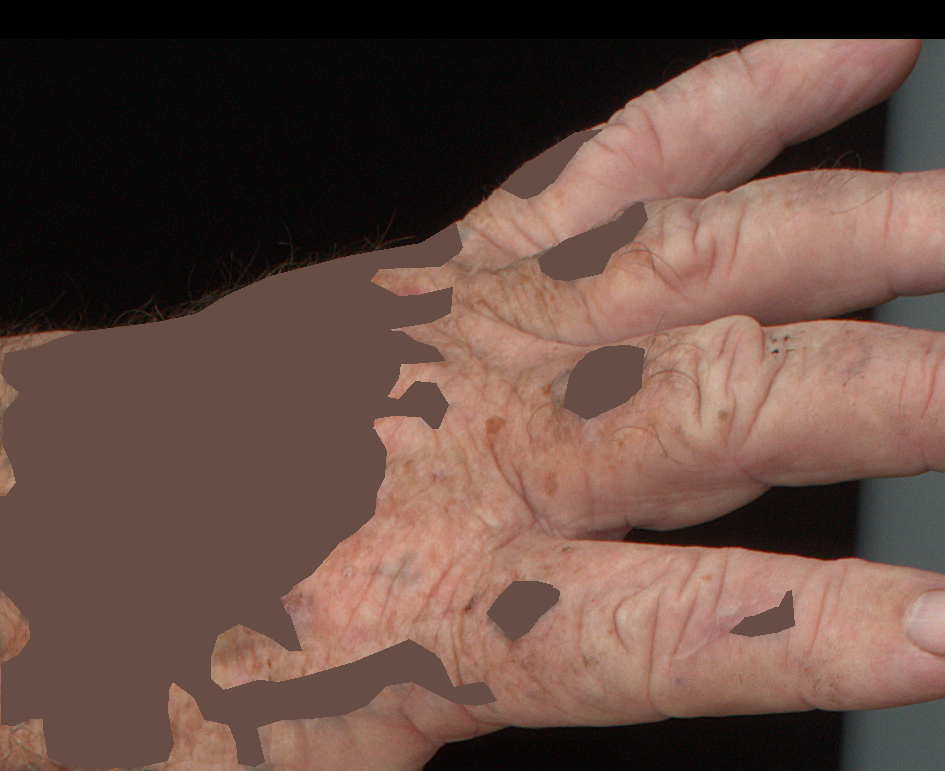}
    \end{minipage}
    \hfill
    \begin{minipage}{0.23\textwidth}
        \centering
        \includegraphics[width=\linewidth]{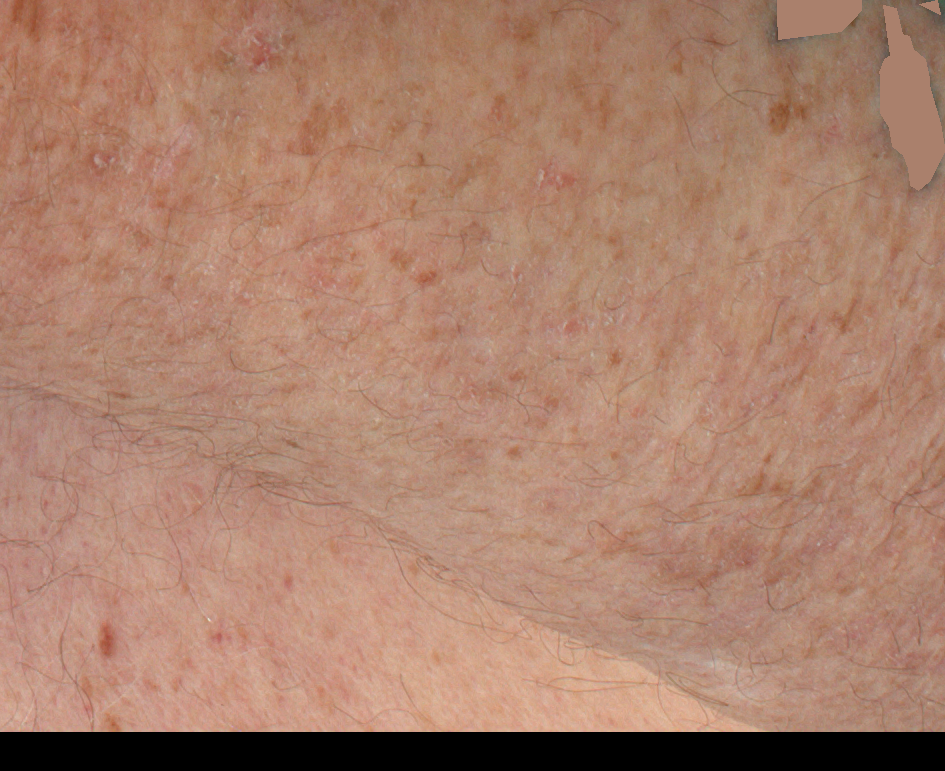}
    \end{minipage}
    \hfill
    \begin{minipage}{0.23\textwidth}
        \centering
        \includegraphics[width=\linewidth]{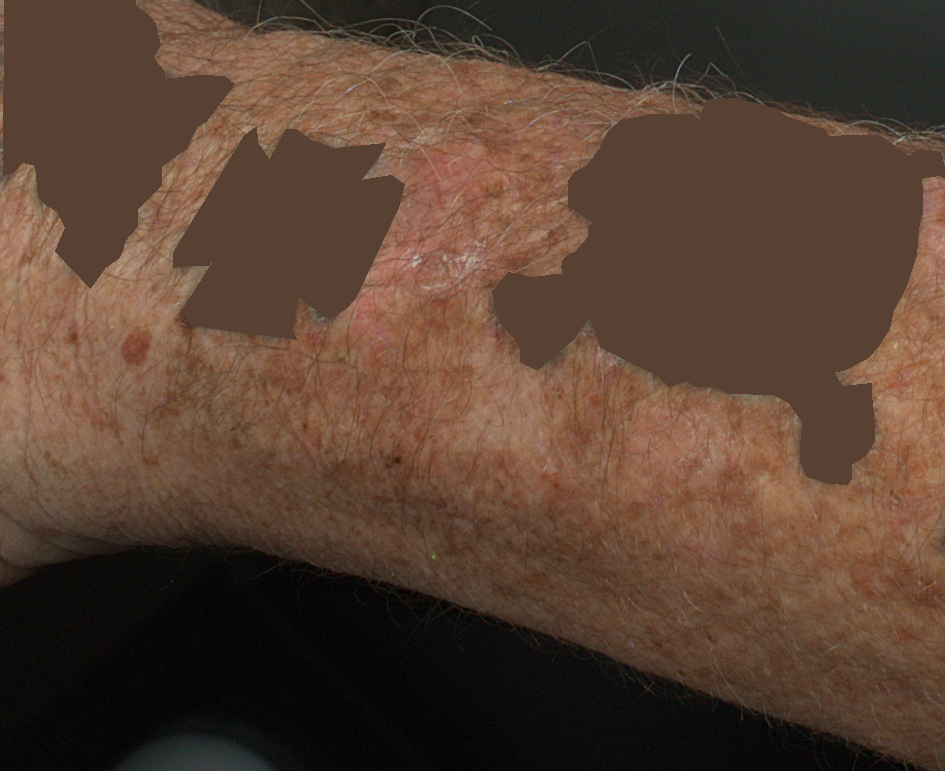}
    \end{minipage}
    \caption{Examples of 2D tiles where tattoos were inpainted during the annotation phase to preserve patient privacy while maintaining skin features for lesion analysis.}
    \label{fig:tattoo-images}
\end{figure}

The annotation workflow, as shown in Figure \ref{fig:v7-workflow}, began with the upload of tiles to the V7 Darwin platform. Within the designed workflow system, \textit{labeller} were assigned batches of images to process. Once completed, the batch of images was automatically routed to the next available \textit{labeller} or \textit{reviewer} depending on the assigned workflow. The \textit{labeller} would assign sun damage and/or tattoo tags, as applicable, and triggered the "Auto Annotate" feature of V7 by creating rough bounding boxes around the lesions. This feature used a pre-trained segmentation model to automatically generate polygons for the lesions, which were subsequently refined to ensure accuracy. While the reviewer would conduct a quality control check for annotation accuracy. An example of masked (inpainting) tattoos is provided in Figure \ref{fig:tattoo-images}, which depicts certain examples of tiles that contain masked tattoos.

\subsection*{Public subset selection}
To select an optimal data subset for public release, we began by analysing potential dataset biases. We identified one extreme category: participants under 30 years of age, which only included two patients. These outlier cases were initially set aside to be strategically distributed throughout the final train and test sets, ensuring that rare but clinically significant cases were preserved in public release. For the remaining tiles, we adapted the \textit{Wallace rule of nines} \cite{wallace1951exposure}, a method traditionally used for assessing burn surface area in dermatology, to establish sampling proportions across anatomical regions. This approach allocated $37.5\%$ of the samples to the torso, $17\%$ to each left and right arms, $13.5\%$ to each left and right legs, and $0.5\%$ to unknown cases, balancing practical considerations with anatomical representation while maintaining clinical relevance.

\begin{figure}[ht!]
    \centering
    \includegraphics[width=\linewidth]{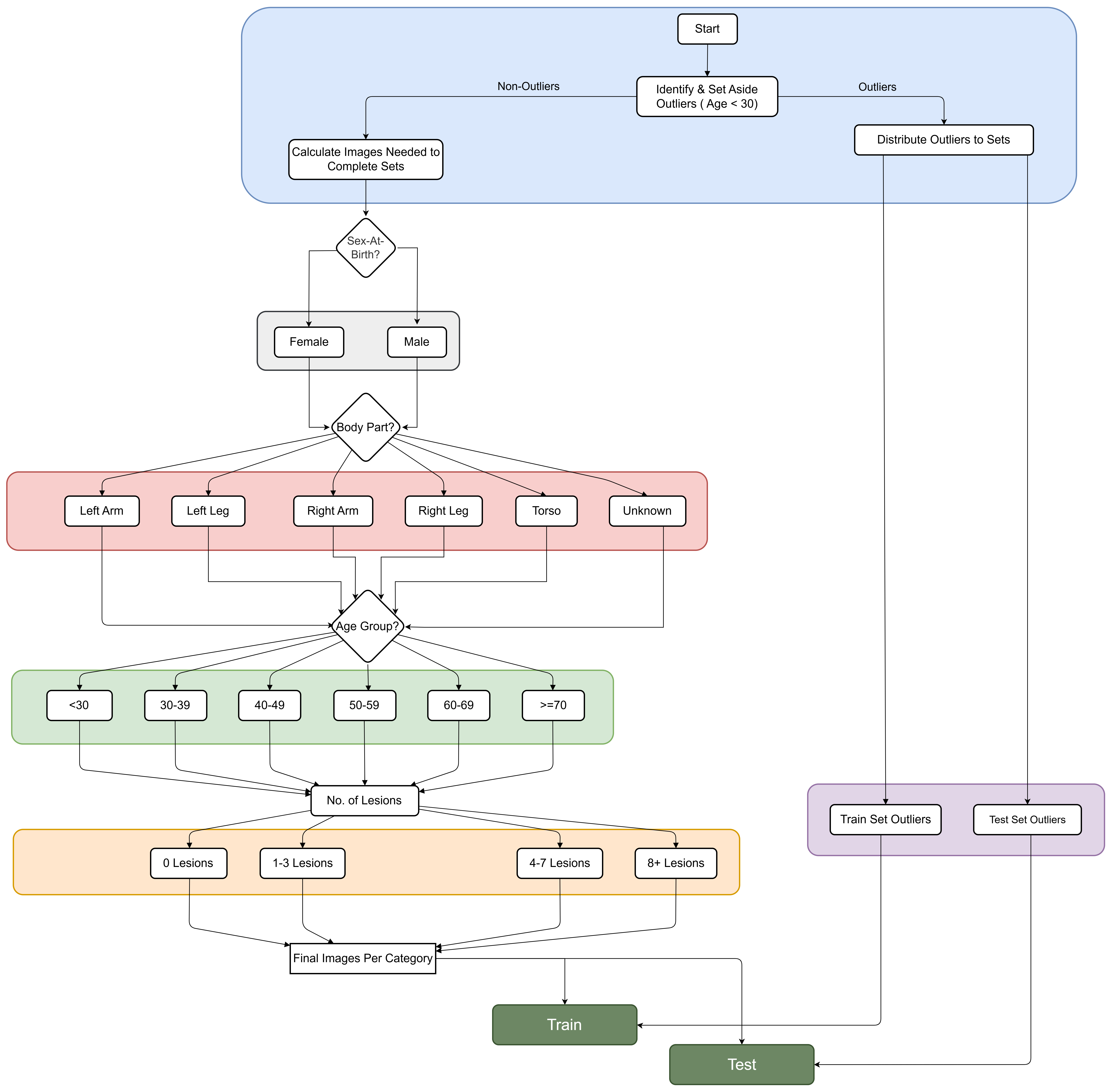}
    \caption{Hierarchical stratified sampling workflow for balanced subset selection. The process begins by setting aside outlier cases (age < 30), then splits the remaining data by sex-at-birth, anatomical region, presence of lesions, and finally applies random sampling within each stratum. The outlier cases are strategically allocated to the train and test sets to ensure representation of rare but clinically significant cases.}
    \label{fig:data-selection-workflow}
    
\end{figure}

The selection process followed a hierarchical stratified sampling approach, as shown in Figure \ref{fig:data-selection-workflow}, to ensure balanced representation across multiple dimensions. This systematic workflow guided the sampling through successive stages of categorisation, starting with sex-at-birth divisions and proceeding through anatomical regions and lesion presence. Within each resulting stratum, random sampling was performed to meet target counts while maintaining a proportional representation. The previously identified outliers were then integrated into the final dataset, ensuring that both common and rare presentations were available for model development and evaluation. 

\subsection*{Acquiring ethical authorisation for data sharing}
This study received the approval of two ethics committees: the Human Research Ethics Committee in Brisbane, Australia, and the Hospital Clinic Barcelona Research Ethics Committee (REC) in Spain. The research was registered with \url{ClinicalTrials.gov} (ref.NCT05955443) and conducted in accordance with Good Clinical Practice guidelines. All study procedures adhered to the ethical principles outlined in the Declaration of Helsinki (1964). Findings from this research will be shared in peer-reviewed journals and presented at international scientific conferences.



\section*{Data Records}
\subsection*{Data Accessibility}
\begin{figure}[!h]
\begin{center}
    \scalebox{0.7}{  
    \begin{forest}
        for tree={
            font=\sffamily,
            folder indent=.1em,
            folder icons=8pt,
            edge=densely dotted,
            grow'=0,
            s sep=1.5pt,
            l sep=8pt
        }
        [dataset
            [train
                [images [...png, is file]]
                [labels [...txt, is file]]
                [metadata.csv, is file]
                [labels.json, is file]]
            [test
                [images [...png, is file]]
                [labels [...txt, is file]]
                [metadata.csv, is file]
                [labels.json, is file]]
        ]
    \end{forest}
    }
\end{center}
\caption{Dataset directory structure illustrating the hierarchical organisation of image tiles, annotations (YOLO and COCO formats), and metadata files across training and test sets of the iToBoS dataset hosted on the Kaggle platform.}
\label{fig:directory-structure}
\end{figure}
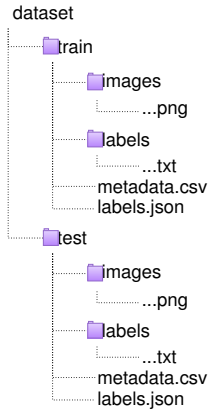

The dataset is available at \url{https://www.kaggle.com/competitions/itobos-2024-detection/data} and is organised in a manner that facilitates easy navigation and usability for research purposes. The dataset is released under a Creative Commons Non-Commercial Attribution (CC-BY NC) license, in accordance with the licensing terms and conditions agreed upon by the providing institutions. It comprises categorised image tiles in PNG format and their corresponding annotations in two formats: text files using YOLO format and JSON files using COCO format. Since this is a single-label dataset, all annotations are assigned the label '0'. Each subset (training and test) is accompanied by a metadata file in CSV format that provides additional information for each image, including anatomical location, patient demographics (age and sex at birth), and sun damage score. The complete organisation of the dataset on Kaggle is illustrated in Figure~\ref{fig:directory-structure}.

However, it should be noted that while the test set images and metadata are currently accessible, the corresponding labels will remain private at least until the completion of the iToBoS lesion detection challenge.

\subsection*{Data Description}

The dataset consists of 8,473 images in the training set and 8,481 images in the test set. In the training set, 6,723 images contain lesions (2,412 from Barcelona and 4,311 from Brisbane) and 1,750 images are without lesions (770 from Barcelona and 980 from Brisbane). Similarly, the test set contains 6,750 images with lesions (3,633 from Barcelona and 3,117 from Brisbane) and 1,731 images without lesions (914 from Barcelona and 817 from Brisbane). To better illustrate the composition of the dataset, we present a comprehensive breakdown of the dataset characteristics in Table~\ref{table:dataset-characteristics}.

\begin{table}[h!]
    \centering
    \renewcommand{\Vline}{\vrule width 0.2pt}
    \resizebox{1.0\textwidth}{!}{%
        \setlength{\tabcolsep}{6pt}
        \renewcommand{\arraystretch}{1.4}
        \footnotesize
        \begin{tabular}{!{\Vline}l!{\Vline}c|c|c!{\Vline}c|c|c!{\Vline}}
            \hline
            \rowcolor{myheader}
            \textbf{Category} & \multicolumn{3}{c|}{\textbf{Train Set}} & \multicolumn{3}{c|}{\textbf{Test Set}} \\
            \cline{2-7}
            &\cellcolor{myblue}Barcelona \footnotesize{(count(\%))} & \cellcolor{mygreen}Brisbane \footnotesize{(count(\%))} & \textbf{Total} & \cellcolor{myblue}Barcelona \footnotesize{(count(\%))} & \cellcolor{mygreen}Brisbane \footnotesize{(count(\%))} & \textbf{Total} \\
            \hline
            \rowcolor{section1}
            \multicolumn{7}{l}{\textbf{Lesion Presence (per Image)}} \\
            With Lesions & 2,412 \footnotesize{(35.9)} & 4,311 \footnotesize{(64.1)} & \textbf{6,723} & 3,633 \footnotesize{(53.8)} & 3,117 \footnotesize{(46.2)} & \textbf{6,750} \\
            Without Lesions & 770 \footnotesize{(44.0)} & 980 \footnotesize{(56.0)} & \textbf{1,750} & 914 \footnotesize{(52.8)} & 817 \footnotesize{(47.2)} & \textbf{1,731} \\
            \hline
            \rowcolor{section2}
            \multicolumn{7}{l}{\textbf{Anatomic Location (per Image)}} \\
            Torso & 1,255 \footnotesize{(39.0)} & 1,966 \footnotesize{(61.0)} & \textbf{3,221} & 1,815 \footnotesize{(56.3)} & 1,410 \footnotesize{(43.7)} & \textbf{3,225} \\
            Left Arm & 486 \footnotesize{(34.1)} & 940 \footnotesize{(65.9)} & \textbf{1,426} & 731 \footnotesize{(49.7)} & 740 \footnotesize{(50.3)} & \textbf{1,471} \\
            Right Arm & 546 \footnotesize{(37.1)} & 927 \footnotesize{(62.9)} & \textbf{1,473} & 840 \footnotesize{(59.1)} & 582 \footnotesize{(40.9)} & \textbf{1,422} \\
            Left Leg & 460 \footnotesize{(38.7)} & 730 \footnotesize{(61.3)} & \textbf{1,190} & 571 \footnotesize{(48.9)} & 597 \footnotesize{(51.1)} & \textbf{1,168} \\
            Right Leg & 435 \footnotesize{(38.0)} & 710 \footnotesize{(62.0)} & \textbf{1,145} & 589 \footnotesize{(51.1)} & 563 \footnotesize{(48.9)} & \textbf{1,152} \\
            Unknown & 0 \footnotesize{(0.0)} & 18 \footnotesize{(100.0)} & \textbf{18} & 1 \footnotesize{(2.3)} & 42 \footnotesize{(97.7)} & \textbf{43} \\
            \hline
            \rowcolor{section3}
            \multicolumn{7}{l}{\textbf{Sex At Birth}} \\
            Male & 9 \footnotesize{(39.1)} & 14 \footnotesize{(60.9)} & \textbf{23} & 14 \footnotesize{(58.3)} & 10 \footnotesize{(41.7)} & \textbf{24} \\
            Female & 13 \footnotesize{(56.5)} & 10 \footnotesize{(43.5)} & \textbf{23} & 11 \footnotesize{(45.8)} & 13 \footnotesize{(54.2)} & \textbf{24} \\
            Unknown & 2 \footnotesize{(66.7)} & 1 \footnotesize{(33.3)} & \textbf{3} & 2 \footnotesize{(66.7)} & 1 \footnotesize{(33.3)} & \textbf{3} \\
            \hline
            \rowcolor{section4}
            \multicolumn{7}{l}{\textbf{Patient Age Groups}} \\
            $<30$ & 0 \footnotesize{(0.0)} & 0 \footnotesize{(0.0)} & \textbf{0} & 2 \footnotesize{(100.0)} & 0 \footnotesize{(0.0)} & \textbf{2} \\
            30-39 & 4 \footnotesize{(66.7)} & 2 \footnotesize{(33.3)} & \textbf{6} & 2 \footnotesize{(50.0)} & 2 \footnotesize{(50.0)} & \textbf{4} \\
            40-49 & 5 \footnotesize{(55.6)} & 4 \footnotesize{(44.4)} & \textbf{9} & 6 \footnotesize{(60.0)} & 4 \footnotesize{(40.0)} & \textbf{10} \\
            50-59 & 8 \footnotesize{(53.3)} & 7 \footnotesize{(46.7)} & \textbf{15} & 9 \footnotesize{(60.0)} & 6 \footnotesize{(40.0)} & \textbf{15} \\
            60-69 & 1 \footnotesize{(10.0)} & 9 \footnotesize{(90.0)} & \textbf{10} & 1 \footnotesize{(12.5)} & 7 \footnotesize{(87.5)} & \textbf{8} \\
            $\geq 70$ & 4 \footnotesize{(66.7)} & 2 \footnotesize{(33.3)} & \textbf{6} & 5 \footnotesize{(55.6)} & 4 \footnotesize{(44.4)} & \textbf{9} \\
            Unknown & 2 \footnotesize{(66.7)} & 1 \footnotesize{(33.3)} & \textbf{3} & 2 \footnotesize{(66.7)} & 1 \footnotesize{(33.3)} & \textbf{3} \\
            \hline
            Total No. of Images & \multicolumn{3}{c|}{\textbf{8,473}} & \multicolumn{3}{c|}{\textbf{8,481}} \\
            Total No. of Annotations & \multicolumn{3}{c|}{\textbf{29,403}} & \multicolumn{3}{c|}{\textbf{30,594}} \\
            \hline
        \end{tabular}
    }
    \caption{Comprehensive Dataset Characteristics: Distribution of lesion presence, anatomical locations, patient demographics (n=49 patients in training, n=51 patients in test), and annotation counts across training and test sets by data collection sites.}
    \label{table:dataset-characteristics}
\end{table}


\noindent As shown in Table~\ref{table:dataset-characteristics}, the dataset maintains varying proportions between Barcelona and Brisbane. The training set exhibits an approximate 4:1 ratio of lesion to non-lesion cases, comprising 6,723 lesion cases and 1,750 non-lesion cases. Similarly, the test set maintains a consistent ratio of approximately 4:1, with 6,750 lesion cases and 1,731 non-lesion cases. The anatomical distribution of the dataset reveals the torso as the primary examination site, accounting for approximately 38\% of images in both the training and test sets. This is followed by a balanced representation of images depicting the arms and legs. Additionally, the sex distribution is notably balanced, with an almost equal proportion of male and female patients in both the training set (23 males and 23 females) and the test set (24 males and 24 females).

\noindent The age distribution reveals geographic variations, with cases under $30$ being exclusively from Barcelona in the test set, while no cases under $30$ exist in the training set. The $60$-$69$ age group shows a strong presence from Brisbane ($90\%$ in training and $87.5\%$ in test set). The middle age groups ($40$-$59$) show a relatively balanced distribution between the two locations. A small portion of cases ($3$ patients in each set) have unknown age and sex information. The training set contains $49$ unique patients with an average of $173$ images per patient, while the test set includes $51$ unique patients with an average of $166$ images per patient. For images containing lesions, there is an average of $4$ lesions per image in the training set and $5$ lesions per image in the test set, as evidenced by the total annotation counts ($29,403$ annotations across $6,723$ lesion images in training, and $30,594$ annotations across $6,750$ lesion images in test). All patients had white skin (Fitzpatrick skin type I-II) and were predominantly of European ancestry.

\subsubsection*{Data Analysis}
We analysed five key aspects of the dataset: (i) spatial distribution of lesions within images, (ii) bounding box properties, (iii) lesion size characteristics (including both categorisation and distribution), (iv) sun damage assessment, and (v) image dimensions.

\begin{figure}[h!]
    \centering
    \includegraphics[width=\linewidth]{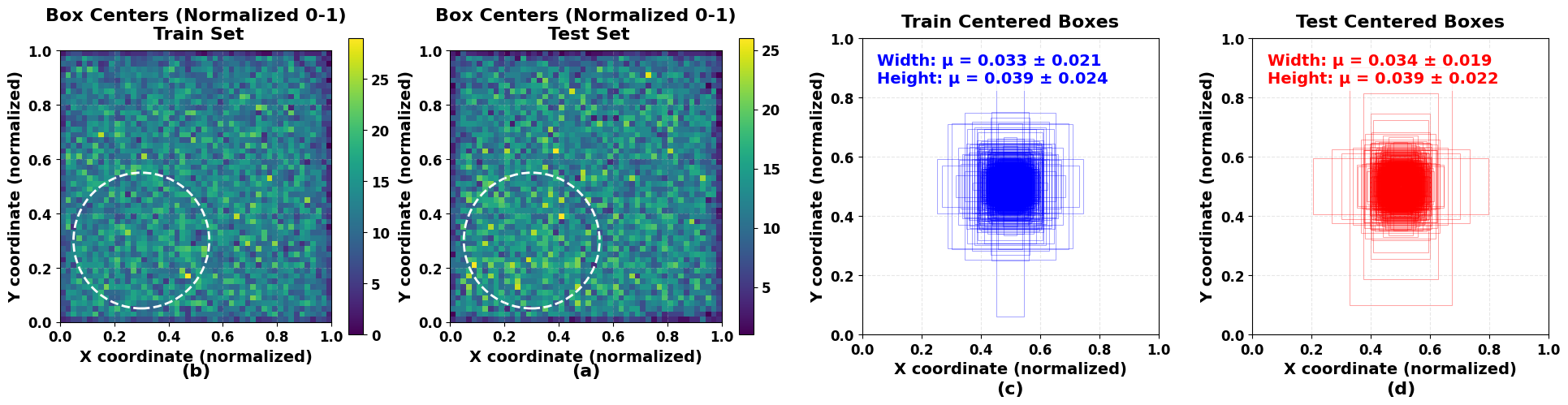}
    \caption{
        Visualization of lesion annotation distributions: 
        (a, b) Spatial distribution heatmaps of lesion centers for the train and test sets, respectively. 
        White dashed circles (\( r = 0.25 \), centered at \( (0.3, 0.3) \)) highlight a region with higher annotation density in the test set. 
        (c, d) Normalized bounding box size comparison with boxes centered at a common origin, showing the train set (blue) with a tighter size distribution 
        (\( \mu \pm \sigma \): width = \( 0.033 \pm 0.021 \), height = \( 0.039 \pm 0.024 \)) compared to the test set (red) 
        (\( \mu \pm \sigma \): width = \( 0.034 \pm 0.019 \), height = \( 0.039 \pm 0.022 \)).
    }
    \label{fig:box_centers}
\end{figure}

The spatial distribution analysis (see Figure~\ref{fig:box_centers} a\&b) examines the normalised locations of lesion centers across all images. To create comparable spatial distributions regardless of original image dimensions, we normalised all bounding box center coordinates by their respective image widths and heights. These normalised coordinates were then used to generate density heatmaps for both training and test sets. The resulting visualisation reveals an even distribution of lesions across the normalised image space. To illustrate the difference in annotation density between sets, we highlight a sample circular region ($r=0.25$, centered at $(0.3,0.3)$, marked by white dashed circle) where, like in other areas of the image, the test set shows higher annotation density compared to the training set, consistent with its larger number of total annotations. For analysing bounding box characteristics (see Figure~\ref{fig:box_centers} c \& d), we collected all boxes in a common center to compare their sizes. The bounding boxes maintain similar dimensions across both sets, with train set showing width $\mu = 0.033 \pm 0.021$, height $\mu = 0.039 \pm 0.024$, and test set showing width $\mu = 0.034 \pm 0.019$, height $\mu = 0.039 \pm 0.022$.

\begin{figure}[h!]
    \centering
    \begin{subfigure}[b]{0.48\textwidth}
        \includegraphics[width=\linewidth]{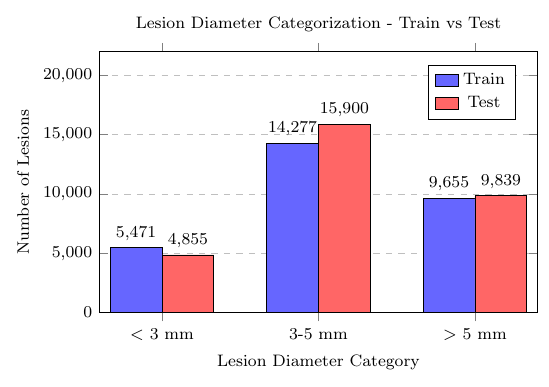}

\caption{Lesion size distribution across clinical categories: small (<3 mm), medium (3–5 mm), and large (>5 mm). The medium-sized lesions dominate both train and test sets, indicating a similar distribution pattern across datasets.}

        \label{fig:diameter-categories}
    \end{subfigure}
    \hfill
    \begin{subfigure}[b]{0.48\textwidth}
        \includegraphics[width=\linewidth]{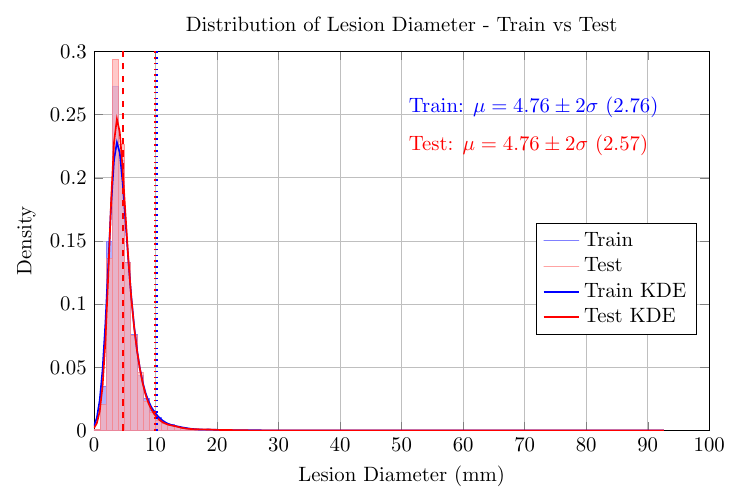}
        \caption{Comparison of lesion diameter distribution between train and test sets, illustrating identical mean values ($\mu=4.76$ mm). The diameter ranges from 0.14 mm to 56.37 mm in the train set and from 0.28 mm to 92.62 mm in the test set.}

        \label{fig:lesion_diameter}
    \end{subfigure}
    \caption{Lesion size analysis showing both clinical categorisation and continuous size distribution characteristics.}
    \label{fig:geometric_chars}
\end{figure}

\begin{figure}[h!]
    \centering
    \begin{subfigure}[b]{0.43\textwidth}
        \includegraphics[width=\linewidth]{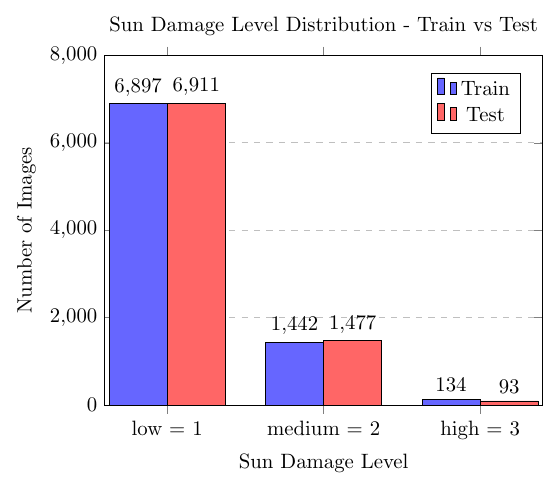}
        \caption{Sun damage severity distribution showing predominance of Level 1 (minimal), followed by Level 2 (moderate), with Level 3 (severe) being rare.}
        \label{fig:sun_damage}
    \end{subfigure}
    \hfill
    \begin{subfigure}[b]{0.55\textwidth}
        \includegraphics[width=\linewidth]{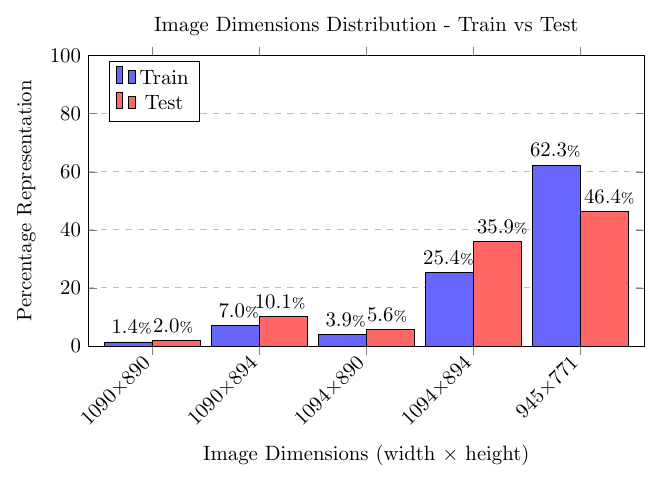}
        \caption{ Image resolution distribution showing the dominance of 
        945 $\times$ 771 px and 1094 $\times$ 894 px formats across the train and test sets, 
        with other formats contributing smaller proportions.}

        \label{fig:dimensions}
    \end{subfigure}
    \caption{Dataset characteristics combining clinical features (sun damage assessment) and technical specifications (image resolutions).}
    \label{fig:clinical_chars}
\end{figure}

Analysing lesion diameters (see Figure~\ref{fig:diameter-categories}), which we calculate as the maximum distance between any two points on the lesion boundary (hull distance) and convert to millimeters using the pixel size information, we grouped the lesions into three clinically relevant categories used by dermatologists in diagnosis: small ($<3$mm), medium ($3$-$5$mm), and large ($>5$mm). We observed that the majority of lesions in both train and test sets fall within the $3$-$5$mm range (train: $14,277$, test: $15,900$), followed by large (train: $9,655$, test: $9,839$) and small categories (train: $5,471$, test: $4,855$). The diameter distribution (see Figure~\ref{fig:lesion_diameter}) exhibits a right-skewed pattern, with a peak around $4$mm and a long tail extending toward larger diameters. The measurements show similar ranges in both sets (train: min=$0.14$mm, max=$56.37$mm; test: min=$0.28$mm, max=$92.62$mm) with identical means (train: $\mu=4.76$, $2\sigma=2.76$ mm; test: $\mu=4.76$, $2\sigma=2.57$ mm).

\noindent The sun damage assessment (see Figure~\ref{fig:sun_damage}) reveals Level 1 (minimal damage) as most prevalent, followed by Level 2 (moderate), and Level 3 (severe) being relatively rare in both sets.

The image dimensions (see Figure~\ref{fig:dimensions}) show predominance of $945\times771$ px (train: $62.3\%$, test: $46.4\%$) and $1094\times894$ px formats (train: $25.4\%$, test: $35.9\%$). Three other dimensions ($1094\times890$ px: train $3.9\%$, test $5.6\%$; $1090\times894$ px: train $7.0\%$, test $10.1\%$; and $1090\times890$ px: train $1.4\%$, test $2.0\%$) comprise the remaining proportions across both splits.
For additional statistical analyses and detailed data distributions, please refer to Appendix~\ref{appendix:dataset_stats}.

\section*{Technical Validation}
All annotated tiles were manually reviewed by a team of dermatologists to ensure the highest level of precision and reliability. These experts carefully evaluated the quality of the annotations carried out by Isahit's \textit{labellers} and \textit{reviewers}, verifying their consistency and accuracy in representing the required details. In cases where discrepancies or inaccuracies were identified, the dermatologists corrected the annotations themselves to align with the expected standards. This rigorous quality control process was essential for maintaining the integrity of the dataset and ensuring that it met the requirements for subsequent analysis and research.

\section*{Usage Notes}
Pixel sizes varied across images due to the different angles and distances between the skin of the patient and the configuration settings of the VECTRA WB360 scanner at both data acquisition sites. When analysing lesion sizes, users are advised to convert pixel measurements using the pixel spacing information provided in the metadata, rather than relying on raw pixel counts, to ensure accurate physical dimensions.
To protect patient privacy, name anonymisation process was implemented to ensure that multiple image tiles from the same patient remain untraceable. The dataset predominantly contains benign lesions, which users may observe when inspecting the types of lesions present in the images. While annotations underwent careful review, users should note there may be occasional inconsistencies. For comprehensive understanding of the dataset characteristics, we direct readers to the statistical analysis in Appendix \ref{appendix:dataset_stats}.

\section*{Code availability}
To support research with this dataset, we provide helper scripts in our GitHub repository (\url{https://github.com/iToBoS/Lesion-Detection-Challange}) for tasks such as data loading, preprocessing, and annotation visualisation. Users seeking additional information should consult the repository documentation or contact the dataset maintainers.

\clearpage
\bibliography{sample}


\section*{Acknowledgements}
This work was funded by the European Union through the iToBoS project (SC1-BHC-06-2020-965221). In addition, we extend our sincere gratitude to the clinical staff at each participating center for their efforts in capturing patient images using the VECTRA WB360 scanner, and to the patients themselves for generously contributing their images to the dataset.

\section*{Author contributions statement}

A. Saha wrote the manuscript and was responsible for data management, collation, and quality assurance. J. Adeola wrote the manuscript, preprocessed the data, curated the annotations, and performed data analysis. J. Malvehy, P. Soyer, and C. Primiero designed the study and collected the data. N. Ferrera and A. Mothershaw performed data extraction and collation, and reviewed the annotations. G. Rezze reviewed and curated the annotations. S. Gaborit was responsible for data annotation. B. D'Alessandro developed the WbTilingTool and provided technical support for the VECTRA WB360 scanner. J. Hudson provided technical support for the V7 Darwin platform. G. Szab\'o and B. Pataki provided cloud infrastructure and technical support for data management. H. Rajani assisted in writing the manuscript, provided technical support, and was responsible for organizing the Kaggle challenge. S. Nazari and H. Hayat assisted in writing the manuscript. R. Garcia conceptualised and supervised the study. All authors reviewed and edited the manuscript.

\section*{Competing interests}

The corresponding author is responsible for providing a \href{https://www.nature.com/sdata/policies/editorial-and-publishing-policies#competing}{competing interests statement} on behalf of all authors of the paper. This statement must be included in the submitted article file.


\clearpage

\section*{APPENDIX}
\appendix

\section{Dataset Statistics}
\label{appendix:dataset_stats}

This appendix presents statistical analyses of the iToBoS dataset. The figures show comparative statistics between train and test sets across multiple characteristics, including demographics, lesion properties, anatomical distribution, and technical aspects.
\begin{figure}[!ht]
   \centering
   \begin{subfigure}[t]{0.48\textwidth}
       \centering
        \includegraphics[width=\linewidth]{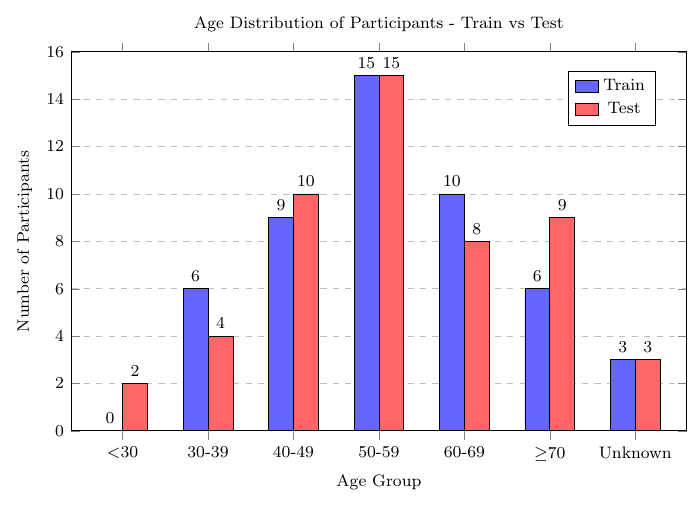}
       
       \caption{Age distribution across train and test sets.}
       \label{fig:age_dist}
   \end{subfigure}
   \hfill
   \begin{subfigure}[t]{0.48\textwidth}
       \centering
        \includegraphics[width=\linewidth]{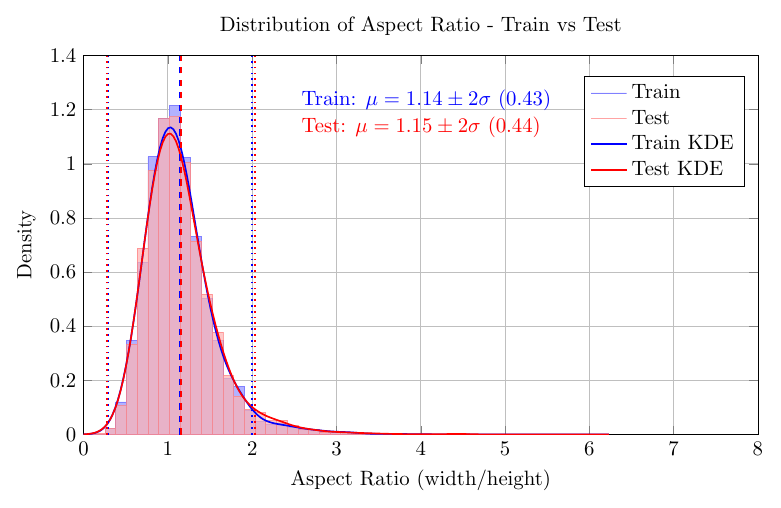}
       \caption{Lesion aspect ratio distribution comparison.}
       \label{fig:aspect_ratio}
   \end{subfigure}
    \caption{Figure 11. Statistical analysis of the iToBoS dataset. (a) Age distribution histogram of 100 participants, with predominant representation in the 50-59 age group (n=15 per test/train) and 3 cases of unknown age in each set; (b) Bounding box aspect ratio distributions between train ($\mu = 1.14 \pm 2\sigma$ (0.43)) and test ($\mu = 1.15 \pm 2\sigma$ (0.44)) sets, demonstrating consistent lesion shape across splits.}
   \label{fig:dataset_stats}
\end{figure}

\end{document}